\documentclass[journal,twocolumn]{IEEEtran}
\usepackage[normalem]{ulem}
\usepackage{blindtext, graphicx}
\usepackage{listings}
\usepackage{graphicx}
\usepackage[font=small]{caption}
\usepackage{color}
\usepackage{amsmath,bm}
\usepackage{amsmath}
\usepackage{amssymb}
\usepackage[ruled,vlined]{algorithm2e}
\usepackage{algpseudocode}
\usepackage{amsthm}
\usepackage{makecell}
\usepackage{siunitx}
\usepackage{comment}
\usepackage{cite}
\usepackage[colorlinks=true, citecolor=blue]{hyperref}
\usepackage{booktabs}
\usepackage{multirow}
\usepackage{mathtools}
\usepackage{booktabs}
\usepackage[utf8]{inputenc}
\makeatletter
\let\NAT@parse\undefined
\makeatother

\newtheorem{proposition}{Proposition}
\newtheorem{definition}{Definition}
\newtheorem{corollary}{Corollary}

\title{Quantifying the Availability of Synchronized and Non-Synchronized Generating Units When Needed}

\author{Yufan Zhang,~\IEEEmembership{Member,~IEEE,}
and Feng Zhao,~\IEEEmembership{Senior Member,~IEEE}%
\thanks{Yufan Zhang and Feng Zhao are with ISO New England Inc.,
Holyoke, MA. The views expressed in this paper are those of the authors and do not represent the views of ISO New England.}
}

\begin{document}

\maketitle

\thispagestyle{empty}
\pagestyle{plain}


\begin{abstract}
Do synchronized units have higher probabilities of being available when needed than non-synchronized units? Power system operation implicitly relies on the qualitative belief that synchronized units are more likely to be available when needed because they are already synchronized to the grid, whereas non-synchronized units must first start and synchronize before becoming available. However, this distinction is rarely expressed through an explicit quantitative measure. To quantify this distinction, we propose failure probabilities for synchronized and non-synchronized generating units, denoted by SynFORd and NonSynFORd, by accounting for their different initial operating states. The complements of the proposed probabilities directly represent the corresponding availability probabilities when the units are needed. Closed-form analytical expressions are derived, revealing the dominant failure mechanisms of the two unit types. Case studies using generating-unit data from the New England system show that non-synchronized units generally exhibit higher and more dispersed failure probabilities than synchronized units.


\end{abstract}

\section{Introduction}

The difference in response reliability between synchronized and non-synchronized generating units is widely recognized in power-system operation, but it is often treated as a qualitative concept rather than an explicit quantitative measure. For example, system operators in the U.S. commonly require a minimum amount of operating reserve to be supplied by synchronized resources, such as the minimum 10-minute spinning-reserve requirement \cite{ISO_NE_OP8,OrtegaKirschen2007}. This reflects the operational belief that synchronized units are more likely to be available when needed because they are already synchronized to the grid and can respond immediately, while non-synchronized units must first start and synchronize before becoming available. However, an explicit quantitative measure for comparing the response reliability of these two types of units is still lacking.

One way to quantify response reliability is through failure probabilities. Well-established indices include forced outage rate (FOR), forced outage rate demand (FORd), and equivalent forced outage rate demand (EFORd) \cite{PJM_Manual22,NERC_GADS}. Specifically, FOR captures full forced outages over the unit's operating exposure, FORd characterizes forced unavailability when the unit is needed, and EFORd extends FORd by incorporating partial forced deratings as equivalent outages. However, these indices generally provide aggregate measures of unit failure performance and do not explicitly distinguish the unit's operating state before it is needed. As a result, they cannot directly quantify the different reliability implications of a synchronized unit that is already in service and a non-synchronized unit that must first start and synchronize.

We address this gap by defining failure probabilities (SynFORd and NonSynFORd) for synchronized and non-synchronized units. SynFORd and NonSynFORd  quantify the conditional probability of a unit being unavailable when needed, while explicitly accounting for its initial operating state. Based on the IEEE four-state model \cite{IEEE_Four_State_Model_1972}, we derive analytical expressions for the proposed failure probabilities using a continuous-time Markov chain (CTMC). We also provide simplified expressions for short response windows, showing that the synchronized-unit failure probability is mainly determined by the in-service failure rate, while the non-synchronized-unit failure probability is mainly determined by the start-failure probability. The proposed formulation provides a quantitative basis for comparing the response reliability of synchronized and non-synchronized units.

\section{State Model and Failure Definitions}

The failure definitions and subsequent derivations are based on a CTMC model. We adopt the well-established IEEE four-state model \cite{IEEE_Four_State_Model_1972} to illustrate the state transitions of a generating unit, as shown in Fig. 1. State 0 represents the reserve-shutdown condition, in which the unit is not synchronized but can be started when needed. State 1 represents a forced-outage condition when the unit is not needed. State 2 represents the in-service condition, in which the unit is synchronized and available. State 3 represents a forced-outage condition during a period of need. Therefore, States 0 and 2 correspond to potentially available conditions, whereas States 1 and 3 correspond to unavailable conditions caused by forced outages. In addition, States 0 and 1 describe conditions in which the unit is not needed, while States 2 and 3 describe conditions in which the unit is needed.
Based on the four-state model, we have the following definition regarding the initial states of synchronized and non-synchronized units.
\begin{figure}[t]
    \centering
    \includegraphics[
        width=0.6\linewidth,
        angle=-90,
        trim={1.3in 0.3in 1.3inin 0.9in},
        clip
    ]{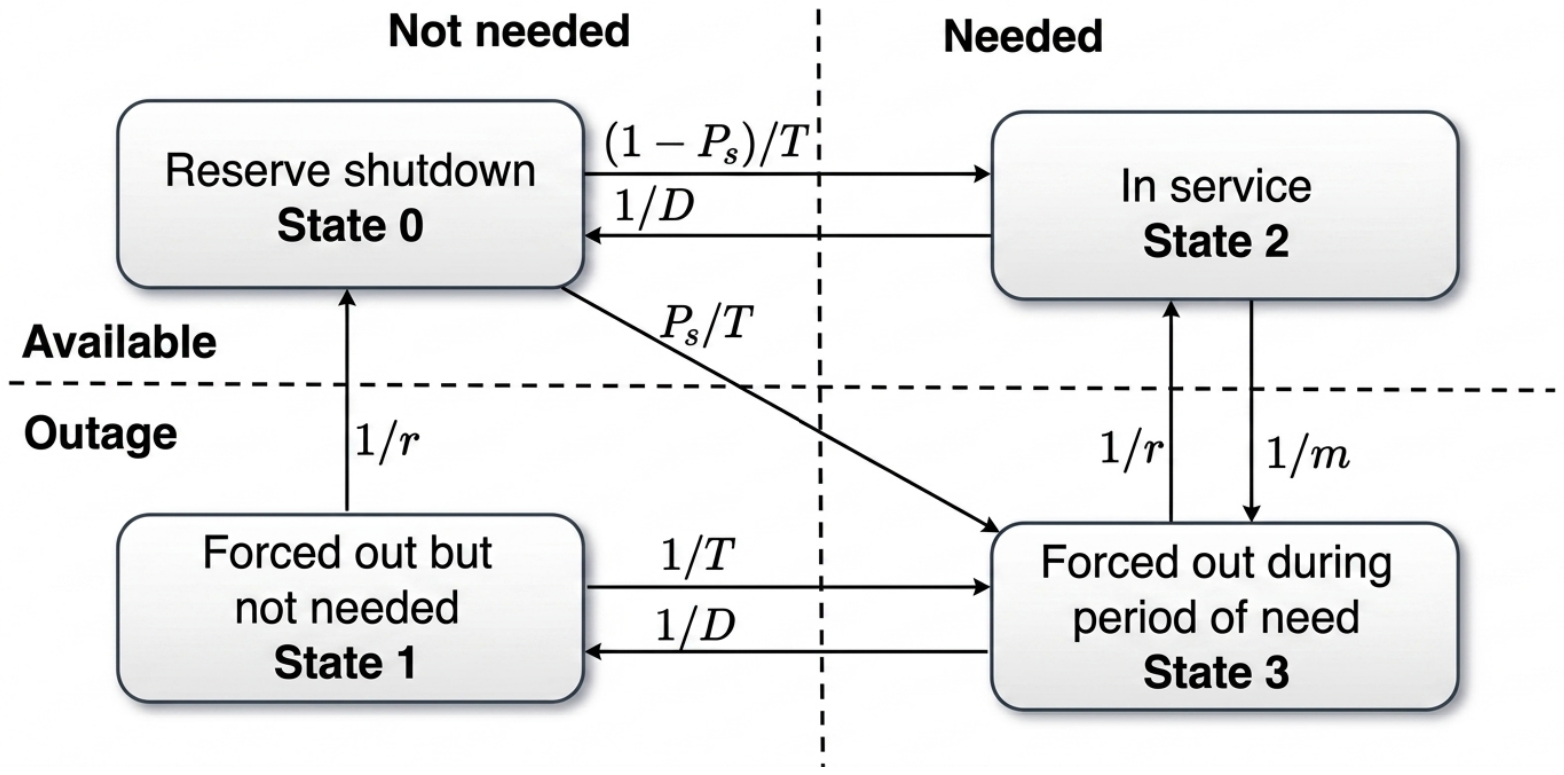}
    \caption{IEEE four-state model. Transition rates are labeled along the corresponding transition arcs; their detailed meanings are provided in \cite{IEEE_Four_State_Model_1972}.}
    \label{fig1}
\end{figure}

\begin{definition}\label{assump1}
    Synchronized and non-synchronized units are initialized in State~2
(in service) and State~0 (reserve shutdown), respectively.
\end{definition}

Consider a response window $[0,t]$, where time 0 denotes the instant at which the unit is needed, we propose the following failure probability definitions of synchronized and non-synchronized units,

\begin{definition}\label{def1}
    SynFORd is the conditional probability of an initially synchronized unit being unavailable when needed at time $t$.
\end{definition}

\begin{definition}\label{def2}
    NonSynFORd is the conditional probability of an initially non-synchronized unit being unavailable when needed at time $t$.
\end{definition}

Let $S_{t_0}$ denote the initial state at time 0 and $S_t$ denote the unit state at time $t$. $\mathcal{O}=\{1,3\}$ denotes the outage states and 
$\mathcal{N}=\{2,3\}$ denotes the needed states. SynFORd and NonSynFORd can then be mathematically formulated as 
\begin{equation}
\label{eq:syn}
\begin{aligned}
\mathbb{P}^{\mathrm{syn}}
&= \mathbb{P}\!\left(S_t \in \mathcal{O} \mid  S_t \in \mathcal{N}, \; S_{t0} = 2\right) \\
&= \mathbb{P}\!\left(S_t = \{1,3\} \mid S_t \in \{2,3\}, \; S_{t0} = 2\right)\\
&= \frac{\mathbb{P}\!\left(S_t = 3, \; S_{t0} = 2\right)}{\mathbb{P}\!\left(S_t \in \{2,3\}, \; S_{t0} = 2\right)}\\
&= \frac{\mathbb{P}\!\left(S_t = 3, \; S_{t0} = 2\right)}{\mathbb{P}\!\left(S_t = 2, \; S_{t0} = 2\right)+\mathbb{P}\!\left(S_t = 3, \; S_{t0} = 2\right)}\\
& = \frac{\mathbb{P}\!\left(S_t = 3\mid  S_{t0} = 2\right)}{\mathbb{P}\!\left(S_t = 2 \mid S_{t0} = 2\right)+\mathbb{P}\!\left(S_t = 3 \mid S_{t0} = 2\right)},
\end{aligned}
\end{equation}

\begin{equation}
\label{eq:non-syn}
\begin{aligned}
\mathbb{P}^{\mathrm{non-syn}}
&= \mathbb{P}\!\left(S_t \in \mathcal{O} \mid  S_t \in \mathcal{N}, \; S_{t0} = 0\right) \\
&= \mathbb{P}\!\left(S_t = \{1,3\} \mid S_t \in \{2,3\}, \; S_{t0} = 0\right)\\
&= \frac{\mathbb{P}\!\left(S_t = 3, \; S_{t0} = 0\right)}{\mathbb{P}\!\left(S_t \in \{2,3\}, \; S_{t0} = 0\right)}\\
&= \frac{\mathbb{P}\!\left(S_t = 3, \; S_{t0} = 0\right)}{\mathbb{P}\!\left(S_t = 2, \; S_{t0} = 0\right)+\mathbb{P}\!\left(S_t = 3, \; S_{t0} = 0\right)}\\
&= \frac{\mathbb{P}\!\left(S_t = 3\mid S_{t0} = 0\right)}{\mathbb{P}\!\left(S_t = 2\mid S_{t0} = 0\right)+\mathbb{P}\!\left(S_t = 3\mid S_{t0} = 0\right)}.
\end{aligned}
\end{equation}


The final equalities in~\eqref{eq:syn} and~\eqref{eq:non-syn} follow from the product rule, since the common factor $\mathbb{P}(S_{t_0}=k)$ appears in both the numerator and denominator and therefore cancels, where $k=2$ for synchronized units and $k=0$ for non-synchronized units.

An appealing property of the proposed definition is that its complement directly represents the probability that a unit is available when it is needed. Specifically, ($1-\mathbb{P}^{\mathrm{syn}}$) and ($1-\mathbb{P}^{\mathrm{non\text{-}syn}}$) give the conditional probabilities that synchronized and non-synchronized units are in State 2, respectively. This interpretation is convenient for power-system reliability applications, where unit availability is often modeled using Bernoulli random variables. Therefore, the proposed failure probabilities can be naturally converted into availability parameters for probabilistic  reliability assessment.

\section{Failure Probability Calculation}

Based on the four-state CTMC model, we next derive analytical expressions for the failure probabilities defined in (1) and (2). Let $M(t)=e^{\Lambda t}$, where $\Lambda$ is the transition-rate matrix of the four-state CTMC, and
$$
\Lambda=\begin{bmatrix}
-\frac{1}{T} & 0 & \frac{1-P_s}{T} & \frac{P_s}{T} \\
\frac{1}{r} & -\left(\frac{1}{r}+\frac{1}{T}\right) & 0 & \frac{1}{T} \\
\frac{1}{D} & 0 & -\left(\frac{1}{D}+\frac{1}{m}\right) & \frac{1}{m} \\
0 & \frac{1}{D} & \frac{1}{r} & -\left(\frac{1}{D}+\frac{1}{r}\right)
\end{bmatrix}
$$
Define entries by state labels. The entry $[M(t)]_{k,j}$ gives the probability that the unit is in State $j$ at time $t$, given that it initially starts from State $k$ at time 0, i.e.,
$$
[M(t)]_{k,j}=\mathbb{P}\left(S_t=j\mid S_{t_0}=k\right).
$$



In the following, we first derive the failure probabilities in \eqref{eq:syn} and \eqref{eq:non-syn} for a general response-window length \(t\). We then present the simplified calculation when \(t\) is small.

\begin{proposition}\label{prop1}
    SynFORd and NonSynFORd in \eqref{eq:syn} and \eqref{eq:non-syn} are calculated as follows,
    \begin{equation}\label{syn2}
    \mathbb{P}^{\mathrm{syn}}=\frac{[e^{\Lambda t}]_{2,3}}{[e^{\Lambda t}]_{2,2}+[e^{\Lambda t}]_{2,3}}
    \end{equation}
    \begin{equation}\label{nonsyn2}
    \mathbb{P}^{\mathrm{non-syn}}=\frac{[e^{\Lambda t}]_{0,3}}{[e^{\Lambda t}]_{0,2}+[e^{\Lambda t}]_{0,3}}
    \end{equation}
where  $\mathbb{P}\!\left(S_t=3 \mid S_{t_0}=2\right)=[e^{\Lambda t}]_{2,3}$, $\mathbb{P}\!\left(S_t=2 \mid S_{t_0}=2\right)=[e^{\Lambda t}]_{2,2}$, $\mathbb{P}\!\left(S_t=3\mid S_{t_0}=0\right)=[e^{\Lambda t}]_{0,3}$, and $\mathbb{P}\!\left(S_t=2 \mid S_{t_0}=0\right)=[e^{\Lambda t}]_{0,2}$.
\end{proposition}

\begin{corollary}\label{col1}
When $t$ is small, the failure probabilities in \eqref{syn2} and \eqref{nonsyn2} can be further simplified as,
\begin{equation}\label{syn3}
    \mathbb{P}^{\mathrm{syn}}\approx t/m
    \end{equation}
\begin{equation}\label{syn3}
    \mathbb{P}^{\mathrm{non-syn}}\approx P_s
    \end{equation}
Note that $P_s$ is the start-failure probability.
\end{corollary}
The proof for Corollary \ref{col1} is provided in the Appendix. The approximation in Corollary \ref{col1} follows from the fact that the probability of two or more state transitions within a sufficiently short response window is negligible. Thus, for synchronized units initially in State 2, the dominant failure event is the direct transition from State 2 to State 3, whose probability is approximately $t/m$. For non-synchronized units initially in State 0, the dominant events are the competing transitions from State 0 to State 2 and from State 0 to State 3. The probability of entering State 3 is therefore $P_s$.

\section{Case Study}

The proposed failure probabilities are evaluated using the NERC Generating Availability Data System (GADS). We study 222 generating units from 2016 to 2025, covering six unit types: combined-cycle (CC), combustion turbine (CT), diesel, hydro, nuclear, and steam units. The response window is set to 15 minutes to match the real-time market clearing interval, so that the calculated probabilities reflect unit availability over the relevant operational decision horizon.

\begin{figure}[t]
    \centering
    \includegraphics[width=0.95\linewidth]{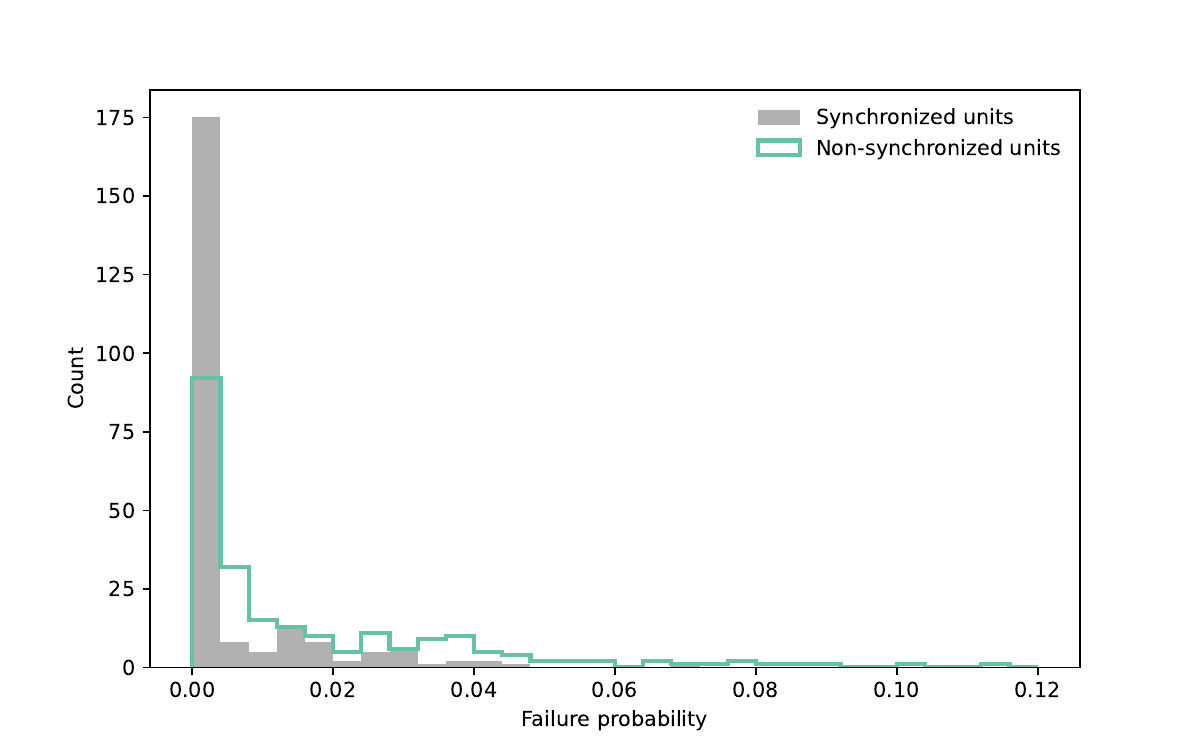}
    \caption{Distributions of the proposed failure probabilities for synchronized and non-synchronized units.}
    \label{fig:prob_dist}
\end{figure}

Fig.~\ref{fig:prob_dist} compares the distributions of the SynFORd and NonSynFORd. SynFORd is highly concentrated near zero, indicating that units already in service have relatively small probabilities of becoming unavailable within the response window. In contrast, NonSynFORd is more dispersed and exhibit a clear right tail, suggesting that start and synchronization-related risks introduce additional uncertainty in its availability.

Table~\ref{tab:prob_by_type} reports the average SynFORd and NonSynFORd by unit type. For most unit types, the average NonSynFORd is higher than the corresponding SynFORd, and this difference is evident for CC, CT, hydro, and steam units. The patterns for diesel and nuclear units are likely influenced by their physical operating characteristics. Diesel units are often operated under specific or infrequent conditions, which may make their estimated synchronized failure probabilities more sensitive to unit-specific outage histories. For nuclear units, start attempts from the reserve-shutdown state were observed very infrequently during the study period, and all observed attempts were successful. Consequently, the empirical estimate of  NonSynFORd is zero.

\begin{table}[t]
\centering
\caption{Average failure probabilities by unit type}
\label{tab:prob_by_type}
\begin{tabular}{lccc}
\toprule
Unit Type & Avg. SynFORd & Avg. NonSynFORd & Count \\
\midrule
CC      & 0.0043 & 0.0047 & 6  \\
CT      & 0.0135 & 0.0263 & 76 \\
Diesel  & 0.0059 & 0.0054 & 13 \\
Hydro   & 0.0004 & 0.0064 & 65 \\
Nuclear & 0.000028 & 0.000000 & 3  \\
Steam   & 0.0008 & 0.0178 & 59 \\
\bottomrule
\end{tabular}
\end{table}

Overall, the results support the operational intuition that non-synchronized units are generally less reliable than synchronized units.

\section{Conclusion}

We propose SynFORd and NonSynFORd to quantitatively compare the response reliability of synchronized and non-synchronized generating units. Using GADS data, the results show that non-synchronized units generally exhibit higher and more dispersed failure probabilities than synchronized units, supporting the operational intuition. A limitation of the current study is that the calculation considers only full forced outages and does not account for partial deratings. Future work will incorporate derating states and quantify their impact on unit availability.

\vspace{-1em}

\bibliographystyle{IEEEtran}
\bibliography{IEEEabrv,mylib}
\vspace{-1em}
\section{Appendix}

When $t$ is sufficiently small, the matrix exponential $e^{\bm{\Lambda}t}$ can be approximated by the first-order Taylor expansion
\begin{equation}
e^{\bm{\Lambda}t}=
\bm{I}+\bm{\Lambda}t+O(t^2).
\end{equation}

For synchronized units, we have,
\begin{align}
\left[e^{\bm{\Lambda}t}\right]_{2,3}
&=
\frac{t}{m}+O(t^2), \
\left[e^{\bm{\Lambda}t}\right]_{2,2}
&=
1-\left(\frac{1}{D}+\frac{1}{m}\right)t+O(t^2).
\end{align}

Substituting these two approximations into \eqref{syn2} gives
\begin{equation}
\mathbb{P}^{\mathrm{syn}}
=
\frac{\frac{t}{m}+O(t^2)}
{1-\left(\frac{1}{D}+\frac{1}{m}\right)t+\frac{t}{m}+O(t^2)} =
\frac{\frac{t}{m}+O(t^2)}{1-\frac{t}{D}+O(t^2)}.
\end{equation}
Hence, when $t$ is small,
\begin{equation}
\mathbb{P}^{\mathrm{syn}}\approx \frac{t}{m}.
\end{equation}

For non-synchronized units, we have,
\begin{align}
\left[e^{\bm{\Lambda}t}\right]_{0,2}
&=
\frac{1-P_s}{T}t+O(t^2), \
\left[e^{\bm{\Lambda}t}\right]_{0,3}
&=
\frac{P_s}{T}t+O(t^2).
\end{align}

Substituting these two approximations into \eqref{nonsyn2} gives
\begin{equation}
\mathbb{P}^{\mathrm{non\text{-}syn}}
=
\frac{\frac{P_s}{T}t+O(t^2)}
{\frac{1-P_s}{T}t+\frac{P_s}{T}t+O(t^2)} 
=
P_s+O(t^2).
\end{equation}
Hence, when $t$ is small,
\begin{equation}
\mathbb{P}^{\mathrm{non\text{-}syn}}\approx P_s.
\end{equation}


\end{document}